\documentclass[twocolumn]{aastex62}

\shorttitle{RR Lyrae Stars in the Field of Sagittarius~II}
\shortauthors{Joo et al.}

\begin{document}

\title{RR LYRAE STARS IN THE FIELD OF SAGITTARIUS~II}

\email{sjjoo250@gmail.com}

\author{Seok-Joo~Joo}
\affiliation{Department of Astronomy and Space Science, Chungnam National University, 99 Daehak-ro, Daejeon 34134, Korea}
\affiliation{Research Institute of Natural Sciences, Chungnam National University, 99 Daehak-ro, Daejeon 34134, Korea}
\affiliation{Korea Astronomy and Space Science Institute, Daejeon 34055, Korea}

\author{Jaemann~Kyeong} \affiliation{Korea Astronomy and Space Science Institute, Daejeon 34055, Korea}
\author{Soung-Chul~Yang} \affiliation{Korea Astronomy and Space Science Institute, Daejeon 34055, Korea}
\author{Sang-Il~Han}
\affiliation{Department of Astronomy and Space Science, Chungnam National University, 99 Daehak-ro, Daejeon 34134, Korea}
\affiliation{Research Institute of Natural Sciences, Chungnam National University, 99 Daehak-ro, Daejeon 34134, Korea}
\affiliation{Korea Astronomy and Space Science Institute, Daejeon 34055, Korea}
\author{Eon-Chang~Sung} \affiliation{Korea Astronomy and Space Science Institute, Daejeon 34055, Korea}
\author{Soo-Chang~Rey} \affiliation{Department of Astronomy and Space Science, Chungnam National University, 99 Daehak-ro, Daejeon 34134, Korea}
\author{Helmut~Jerjen} \affiliation{Research School of Astronomy and Astrophysics, Australian National University, Canberra, ACT 2611, Australia}
\author{Hak-Sub~Kim} \affiliation{Korea Astronomy and Space Science Institute, Daejeon 34055, Korea}
\author{Dongwon~Kim} \affiliation{Department of Astronomy, University of California, Berkeley, CA 94720-3411, USA}
\author{Hyunjin~Jeong} \affiliation{Korea Astronomy and Space Science Institute, Daejeon 34055, Korea}
\author{Chang~H.~Ree} \affiliation{Korea Astronomy and Space Science Institute, Daejeon 34055, Korea}
\author{Sang-Mok~Cha}
\affiliation{Korea Astronomy and Space Science Institute, Daejeon 34055, Korea}
\affiliation{School of Space Research, Kyung Hee University, Yongin, Kyeonggi 17104, Korea}
\author{Yongseok~Lee}
\affiliation{Korea Astronomy and Space Science Institute, Daejeon 34055, Korea}
\affiliation{School of Space Research, Kyung Hee University, Yongin, Kyeonggi 17104, Korea}

\begin{abstract}

We present the detection of RR Lyrae variable stars in the field of the Sagittarius~II (Sgr~II) ultra-faint dwarf (UFD) galaxy.
Using $B$, $V$ time-series photometry obtained with the Korea Microlensing Telescope Network (KMTNet) 1.6\,m telescope at CTIO
and $G$-band data from $Gaia$ Data Release 2 (DR2), we identified and characterized two $ab\,$-type and four $c\,$-type RR Lyrae variables.
Five out of the six stars are clustered within three half-light radii ($\sim$$4\farcm8$) of the galaxy indicating their association with Sgr~II,
while the RRab star V4 is located $\sim$22$\,'$ from the galaxy center.
By excluding V4, the high $c\,$-type fraction (0.8) and the period of the only RRab star V3 ($P_{V3} \simeq 0.666$~days)
suggest an Oosterhoff~II (Oo~II) classification for Sgr~II.
Located close to the locus of Oo~II clusters in the period-amplitude diagram, V3 is similar to RRab stars
in other UFDs having Oosterhoff-intermediate and Oo~II properties.
Sgr~II is, however, more compact than usual UFDs, placed in between star clusters and dwarf galaxies in the size-luminosity plane,
and therefore spectroscopic studies are eventually required to ascertain the true nature of this stellar system.
We derive the metallicity ($\rm[Fe/H]_{RRab} \simeq -2.1 \,\pm\, 0.3$) and heliocentric distance ($\sim$$64\pm3$\,kpc) of Sgr~II
from the RR Lyrae stars, and estimate its age ($\sim$12\,Gyr) based on our stellar population models.
The Oosterhoff properties of UFDs can be explained with the evolution effect of RR Lyrae stars in the instability strip.

\end{abstract}

\keywords{galaxies: dwarf --- galaxies: individual (Sagittarius~II) --- Local Group --- stars: variables: RR Lyrae}

\defcitealias{joo18}{Paper~I}

\section{Introduction} \label{sec:intro}

Since 2005, about 40 dwarf satellite galaxies have been discovered around the Milky Way (MW) from large optical surveys such as SDSS \citep{yor00}, DES \citep{des16},
and Pan-STARRS1 \citep{cha16}, greatly exceeding the number of twelve classically known dwarf satellites.
Most of the new stellar systems are ultra-faint dwarfs (UFDs), the smallest and least luminous galaxies known.
Expected to be completely dark matter dominated \citep[e.g.][]{sim07}, detailed investigations of these UFDs are of particular importance
to better understand the formation and evolution of galaxies in the current $\Lambda$CDM hierarchical merging paradigm \citep[e.g.,][]{kau93}.

RR Lyrae stars provide an excellent opportunity to study resolved stellar populations. They are well-known standard candles and
tracers of old, metal-poor populations. Metallicity and line-of-sight reddening can also be derived from their variability characteristics.
By comparing them with those in other systems including the MW halo, we can place constraints on the formation and evolution of the MW and dwarf galaxies \citep[e.g.,][]{cle10}.
Thus far, about 15 UFDs with total luminosities $M_V > -7$ have been studied for RR Lyrae stars,
and it was reported that all of them have at least one RR Lyrae star \citep[e.g.,][]{viv16},
with the only exception of Carina~III \citep{tor18}.

In this paper, we investigate RR Lyrae stars in the field of the Sagittarius~II (Sgr~II) dwarf galaxy,
as part of our ongoing time-series study of dwarf satellites in the southern hemisphere
using the Korea Microlensing Telescope Network (KMTNet) 1.6\,m telescope at Cerro Tololo Inter-American Observatory (CTIO).
Recently discovered by \citet{lae15} from the Pan-STARRS1 survey, Sgr~II has a total luminosity
of $M_{V} \simeq -5.2$ and a heliocentric distance of $\sim$67\,kpc \citep[see also][]{mut18}.
Compared to typical UFDs with similar luminosities, it is relatively small and compact (half-light radius, $r_{h} \approx 32$\,pc or $1\farcm 6$),
but is larger than usual globular clusters (GCs).
In the size-luminosity plane, Sgr~II can be classified as either the most compact UFD or one of the most extended GCs.
For that reason, \citet{lae15} also assigned an alternative name (Laevens~5) for this stellar system.
We identify and analyze RR Lyrae variables by fitting light curve templates to our time-series photometry.
The data from $Gaia$ Data Release 2 (DR2)\citep{gai16,gai18} is also used to complement our data.
We compare them with those stars of other UFDs in the period-amplitude diagram and discuss the implications of our results.

\section{Observations, Data Reduction, and Color-Magnitude Diagrams} \label{sec:obs}

Time-series observations of the field of Sgr~II were conducted in the $B$- and $V$-bands using the KMTNet-CTIO 1.6\,m telescope on 17 nights
from October to November in 2016 and from September to October in 2017.
In total, 117 and 121 frames were obtained for $B$- and $V$-bands, respectively, with an exposure time of 120\,s each.
Point-spread function (PSF) photometry was performed using DAOPHOT~II/ALLSTAR \citep{ste87} and ALLFRAME \citep{ste94}
on the time-series data preprocessed by the KMTNet pipeline \citep{kim16}.
We used the AAVSO Photometric All-Sky Survey (APASS) database (\url{http://www.aavso.org/apass}) for standardization
and SCAMP \citep{ber06} for astrometry by adopting the third US Naval Observatory (USNO) CCD Astrograph Catalog, UCAC3 \citep{zac10}.
We also used the time-series data from $Gaia$ DR2 \citep{gai18} for stars identified as variables in the field of Sgr~II.

In the left and middle panels of Figure~\ref{fig1}, we present color-magnitude diagrams (CMDs) of the Sgr~II field,
for the region inside $r_{h}$ ($\sim$$1\farcm 6$; \citealt{mut18}) and 3\,$r_{h}$ ($\sim$$4\farcm 8$), respectively.
To obtain the CMDs, we combined our time-series images with SWarp \citep{ber02} and
carried out PSF photometry with DAOPHOT~II/ALLSTAR \citep{ste87}.
In both panels, blue horizontal branch (HB) and red giant branch (RGB) stars of Sgr~II are prominent over the background of field stars.
The CMD features of Sgr~II are comparable to those presented in \citet[see their Figure~1]{lae15} and \citet[see their Figure~1]{mut18},
while the CMD in the latter is much deeper than ours.
RR Lyrae stars identified in this study are also plotted as crosses in the panels (see next section).
In the right panel, we compare our population models with the foreground-subtracted Hess diagram for 2\,$r_{h}$ ($r< 3\farcm 2$) region,
which will be discussed in Section~\ref{sec:discussion}.

\section{RR Lyrae Stars in the Field of Sgr~II} \label{sec:rrl}

RR Lyrae variables in the field of Sgr~II were detected and characterized using the template light curve fitting technique
outlined by \citet{yan12} and \citet{yan14}.
We selected variable candidates by examining the variability of stars at the level of the HB,
and analyzed them by fitting template light curves.
Readers are referred to \citet[hereafter \citetalias{joo18}]{joo18} for details of the procedure we employed.
Finally, six RR Lyrae (two RRab and four RRc) stars were discovered in the Sgr~II field.
We found that five of the six variables are also identified as RR Lyrae stars in $Gaia$ DR2 \citep{gai18}.
For these five stars, V2 to V6, we included the $Gaia$ time-series data in our light curve analysis
by transforming the $Gaia$ $G$-band to $V$-band magnitudes based on the relations in Table~A2 of \citet[][see also \citealt{cle19}]{eva18}.
The light curves obtained with our best-fit templates are presented in Figure~\ref{fig2}, where the green points are
the $Gaia$ $G$-band data converted to $V$-band. In Table~\ref{tbl1}, we summarize positions and pulsation properties of RR Lyrae stars
including variable type, period, intensity-weighted mean magnitude, the number of observations, and amplitude.
We also list the information (ID, type, period, and the number of data points) from $Gaia$ DR2 for the five variables matched with our data.
Note that the type and period values from $Gaia$ DR2 are almost identical to those estimated in our analysis.

The left panel of Figure~\ref{fig3} is the CMD showing a close-up view of the HB region for stars within $4\farcm 8$ from the galaxy center.
The six RR Lyrae stars detected in this study are well located in the empirical instability strip,
which is estimated from the data of nine Galactic and Large Magellanic Cloud clusters in \citet[][Table 7 in that paper]{wal98}.
We see that V1 is classified as $c\,$-type from the light-curve shape and short period, while it is the reddest RR Lyrae in the CMD.
This star appears to be slightly affected by bleeding from a saturated star, roughly up to $\sim$0.2\,mag more in $V$-band than $B$-band.
If this effect is taken into account, the $B-V$ color of V1 would be similar to those of the other $c\,$-type variables.

The middle panel of Figure~\ref{fig3} presents the spatial distribution of the RR Lyrae and blue HB stars,
while the right panel displays that of all point sources detected in the field for comparison.
The RR Lyrae stars are clearly clustered around the galaxy center estimated by \citet{mut18}. Five out of the six RR Lyrae stars
are within 3$\,r_h$ (i.e., $r< 4\farcm 8$), while the RRab star V4 is located $\sim$22$\,'$ away from the center.
The blue HB stars, that we define as stars in the color and magnitude ranges of $0.0 < B-V < 0.27$ and $19.5 < V < 20.3$, are also centrally concentrated.
We do not find, however, such a clustering of stars belonging to the red HB region, i.e., $0.50 < B-V < 0.65$ at the same magnitude range.
This suggests that there are only a few red HB stars in Sgr~II, if any, indicating the blue HB morphology of Sgr~II in conjunction with the CMDs in Figure~\ref{fig1}.

The positions in the CMD and the spatial distribution of the RR Lyrae and blue HB stars
strongly indicate that most of these stars near the center are associated with Sgr~II.
For the stars outside $\sim$6$\,r_h$ (i.e., $r \gtrsim 9\farcm 6 $) including V4, it is natural to consider them as field stars,
given the heliocentric distance to the galaxy ($\sim$64\,kpc, see below).
By excluding the $ab\,$-type star V4, the fraction of $c\,$-type RR Lyrae variables is calculated to be, $N(c)/N(ab+c) = \frac{4}{5}=0.8$.
This high fraction of RRc stars is also consistent with the blue HB morphology of the system.
The period of the only $ab\,$-type star in Sgr~II, V3, is $P_{ab} \simeq 0.666$~days,
and the mean period of the four $c\,$-type stars is $\langle P_c \rangle = 0.337 \pm 0.023$~days, where the uncertainty is the standard error of the mean.
Note that the long period of the RRab star V3 and the high $c\,$-type fraction suggest an Oosterhoff~II (Oo~II) classification\footnote{
Note however that the Oosterhoff classification is basically based on the mean period of RRab stars and therefore usually applied
to the systems hosting at least five RRab stars \citep[see, e.g.,][]{cat09}. \citet{bra16} have further shown that, using Galactic GCs with more than
35 RR Lyrae stars and nearby dwarf galaxies, the difference in the mean RR Lyrae periods between the Oosterhoff groups (i.e., Oosterhoff dichotomy)
can be explained by changes in metallicity. \label{footnote1}}
for Sgr~II \citep[see, e.g.,][]{cat09}. The mean apparent $V$ magnitude of the RR Lyrae stars is $\langle V_{\rm RR} \rangle = 19.77 \pm 0.04$~mag,
where we also exclude V1, which is affected by bleeding, as well as V4, and the uncertainty is the standard error of the mean.

Figure~\ref{fig4} shows the period-amplitude diagram of the RR Lyrae stars in the Sgr~II field, and compares them
with those in the MW halo and 13 UFDs\footnote{Segue~1 is not included here because of the absence of accurate period and amplitude information,
while \citet{sim11} has reported one or two RR Lyrae stars \citep[see also][]{viv16}.}
with $M_V > -7$, as an update of Figure~10 in \citet{viv16} and Figure~6 in \citetalias{joo18}.
The solid and dotted lines represent the mean distribution of RR Lyrae stars in the MW GCs that belong to the Oosterhoff~I (Oo~I) and Oo~II groups, respectively,
adopted from \citet{zor10} and \citet{cac05}.
It is clear from Figure~\ref{fig4} that RRab stars in the UFDs studied so far have mostly Oosterhoff-intermediate (Oo-int) or Oo~II properties \citep[see also][]{cle10,viv16},
in contrast to those in the MW halo where the majority ($\sim$73$\%$) has Oo~I properties \citep{zin14}.
Placed close to the locus of Oo~II clusters, V3 is in the middle of the distribution of stars in the UFDs in terms of both period and amplitude.\footnote{
V3 is also consistent with the findings that faint dwarf spheroidal galaxies and UFDs show a lack of high-amplitude short-period RRab stars \citep[e.g.,][]{fio17}.}
V4 is, however, almost on the Oo~I line, quite different from most RRab stars in the UFDs.
This supports the previous assumption based on its spatial location that V4 is not a member of Sgr~II but belongs to the MW halo.
We can then conclude that Sgr~II is also similar to the other UFDs in the distribution of RRab stars on the period-amplitude diagram
and is classified as Oo~II.\textsuperscript{\ref{footnote1}}

The blue HB morphology of Sgr~II and the similarity in RR Lyrae properties between Sgr~II and other UFDs suggest that
Sgr~II can be considered as a UFD, because outer halo GCs ($\gtrsim$40\,kpc) tend to have red HB morphologies \citep[e.g,][]{lee94,lae15}.
However, this does not mean that Sgr~II cannot be a GC. There are GCs with Oo~II characteristics similar to Sgr~II in the inner halo \citep{cle01},
and \citet{mut18} has further shown that its structural parameters are more consistent with extended GCs.
Spectroscopic studies are eventually required to ascertain whether it is a galaxy or a star cluster,
by investigating velocity and metallicity dispersions of the system and/or a possible presence of Na-O or Mg-Al anti-correlations shown only in GCs.

Using the properties of the RRab star V3, we derive the interstellar reddening in the direction and metallicity of Sgr~II.
From the empirical period-amplitude-metallicity relation by \citet[][see equations (1) and (2) in that paper]{alc00},
$\rm[Fe/H] = -2.1 \,\pm\, 0.3$ is obtained, where the uncertainty reflects the intrinsic accuracy of the relation.
This is in good agreement with the estimate ([Fe/H]$\approx$$-2.2$) from isochrone fitting by \citet{lae15}.
From Sturch's method \citep{stu66} revised by \citet[][see equation (2) in that paper]{wal90} and the (period-)amplitude-color-metallicity relation by
\citet[][see equations (4) and (5) in \citetalias{joo18}]{pie02}, we infer $E(B-V) \approx 0.18$ and 0.15, respectively.
These $E(B-V)$ values are, however, somewhat larger than that of \citet{lae15}, $\sim$0.097, measured from the dust maps of \citet{sch98} and \citet{sch11}.
Because the reddening value from RRab stars strongly depends on the observed color (or color distribution) of them,
we have less confidence in our reddening estimation based on a single RRab star.
Instead, by adopting $E(B-V) = 0.097$ to the instability strip, we obtain a reasonable match with the observed mean colors of the RR Lyrae stars
in the CMD (see Figure~\ref{fig3}a).

We also estimate the distance to Sgr~II using optical Period-Wesenheit (PW) relations provided by \citet{mar15}, which have
the key advantage that they are reddening free and only marginally dependent on metallicity \citep[see also][for a recent application of the PW relations]{bon19}.
In practice, we adopt metal-independent $(V,B-V)$ PW relations \citep[see their equation~(10) and Table~9]{mar15} for the four RR Lyrae stars (excluding V1 and V4),
and obtain the mean distance modulus, $(m-M)_0 = 19.03 \pm 0.10$\,mag, and the mean heliocentric distance, $\rm d_{\odot} = 64 \pm 3$\,kpc, for Sgr~II.
The uncertainties are standard deviations.
Our distance estimate agrees with those presented by \citet[$67 \pm 5$\,kpc]{lae15} and \citet[$70.2 \pm 5.0$\,kpc]{mut18} to within the uncertainty.
In the case of the probable field RRab star V4, a distance modulus of $\sim$18.58\,mag and a distance of $\sim$52\,kpc are obtained with the same relations,
which implies that V4 is $\sim$12\,kpc closer than the center of Sgr~II.

\section{DISCUSSION} \label{sec:discussion}

\begin{figure*}
\epsscale{1.0}
\plotone{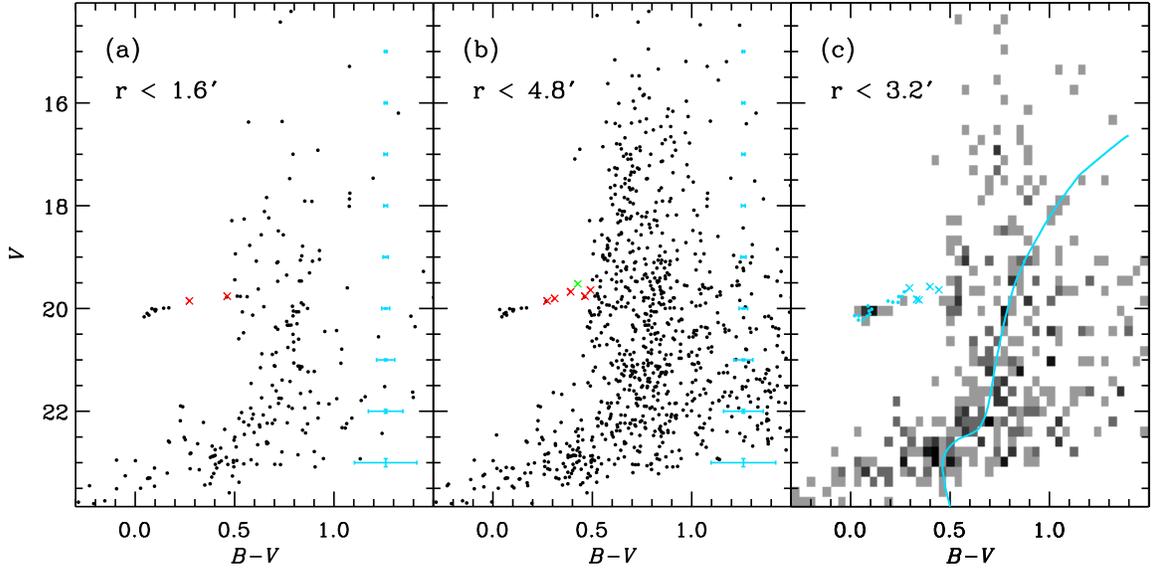}
\caption{
(a) and (b) CMD of stars within $1\farcm 6$ ($r_{h}$) and $4\farcm 8$ (3$\,r_{h}$) regions in the field of Sgr~II, respectively.
Detected RR Lyrae variables are highlighted as crosses, while the green cross denotes V4, which is $\sim$22$\,'$ from the center.
The cyan error bars represent the measurement errors.
(c) Our population models (cyan line and points) superimposed on the foreground-subtracted Hess diagram for $3\farcm 2$ (2\,$r_{h}$) region.
The foreground field was assumed to be an annulus of $9\farcm 6~(6\,r_{h}) < r < 13\farcm 9$ (covering 10 times the area of 2\,$r_{h}$) and normalized.
Model RR Lyrae stars are presented as cyan crosses. Adopted parameters are $t=12\,$Gyr, $\rm[Fe/H]=-2.1$, $\rm[\alpha/Fe] = 0.3$, $E(B-V) = 0.097$,
and $(m-M)_0 = 19.03$\,mag, respectively.
\label{fig1}}
\end{figure*}

\begin{figure*}
\epsscale{1.0}
\plotone{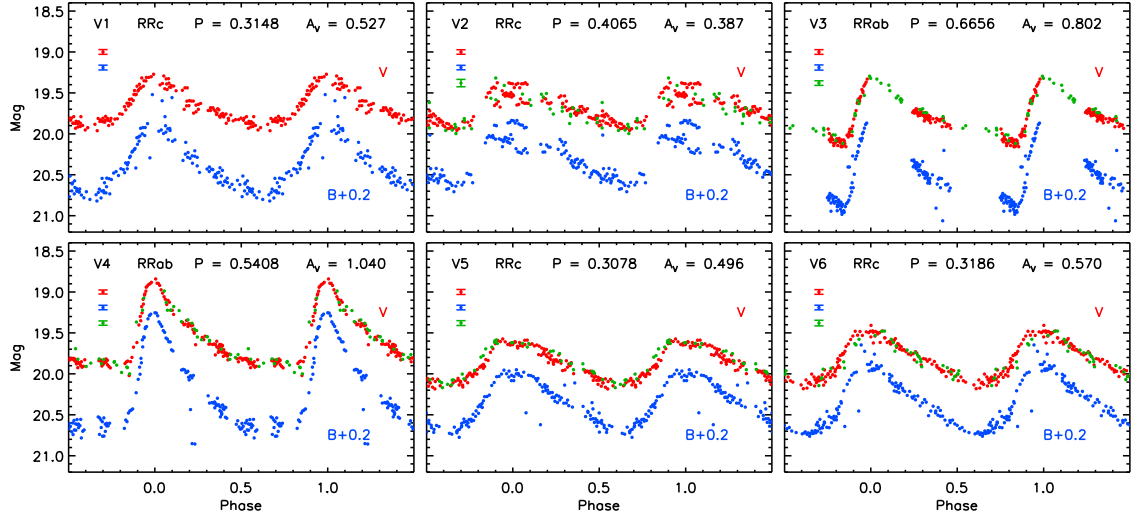}
\caption{
Light curves of the six RR Lyrae stars for $B$- and $V$-bands (blue and red points), respectively.
The $B$-band light curves are shifted by +0.2\,mag. Green points are the $Gaia~G$-band data transformed to $V$-band magnitudes.
The variable information including ID, type, period (days), and $V$-band amplitude (mag) is also presented in each panel.
The error bars in the upper left corner are the mean photometric errors for each band.
The $G$-band uncertainties are estimated using the relation in \citet[see their footnote~6]{hol18}.
\label{fig2}}
\end{figure*}

\begin{figure*}
\epsscale{1.0}
\plotone{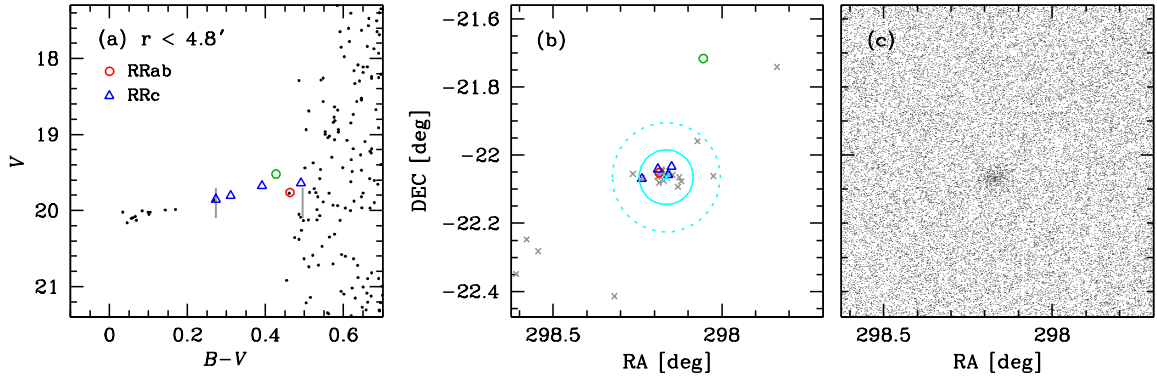}
\caption{
(a) CMD of stars within $4\farcm 8$ (3$\,r_{h}$) zoomed around the HB region, (b) spatial distribution of the RR Lyrae (small circles and triangles)
and blue HB (gray crosses) stars, and (c) same as (b) but for all point sources detected in the field.
The blue HB stars are defined as stars in the ranges of $0.0 < B-V < 0.27$ and $19.5 < V < 20.3$.
The two gray vertical lines in panel~(a) represent the empirical instability strip from \citet{wal98}, which is reddened by 0.097\,mag (see the text).
The solid and dotted cyan circles in panel~(b) indicate 3$\,r_{h}$ ($4\farcm 8$) and 6$\,r_{h}$ ($9\farcm 6$) regions, respectively.
The RR Lyrae and blue HB stars are clearly clustered around the galaxy center (cyan cross) from \citet{mut18}.
It is natural to consider V4 (small green circle) as belonging to the MW halo.
\label{fig3}}
\end{figure*}

To estimate the age of the stellar population in Sgr~II, we have constructed population models using Yonsei-Yale (Y$^2$) isochrones
and HB evolutionary tracks \citep{yi08,han09}, based on the techniques developed by \citet{lee90,lee94} and \citet{joo13}.
With the fixed metallicity of $\rm[Fe/H] = -2.1$ and alpha-element enhancement of $\rm[\alpha/Fe] = 0.3$,
the age value was adjusted until the model best matches the observed CMD.
Our best-fit model with an age of 12 Gyr is presented in the right panel of Figure~\ref{fig1}.
Given the large observational errors near the main sequence turn-off (MSTO), the uncertainty in this age determination can be relatively large,\footnote{
The quadratic sum of the errors in the $V$-band magnitude and extinction (i.e., reddening) is $\sim$0.12\,mag at the MSTO,
which means that the age uncertainty caused by the photometric errors could be more than 1\,Gyr.}
but an old age is clearly required to reproduce the blue HB morphology of Sgr~II, which is comparable to those of other UFDs \citep[e.g.,][]{bro14}.

\begin{figure*}
\epsscale{0.8}
\plotone{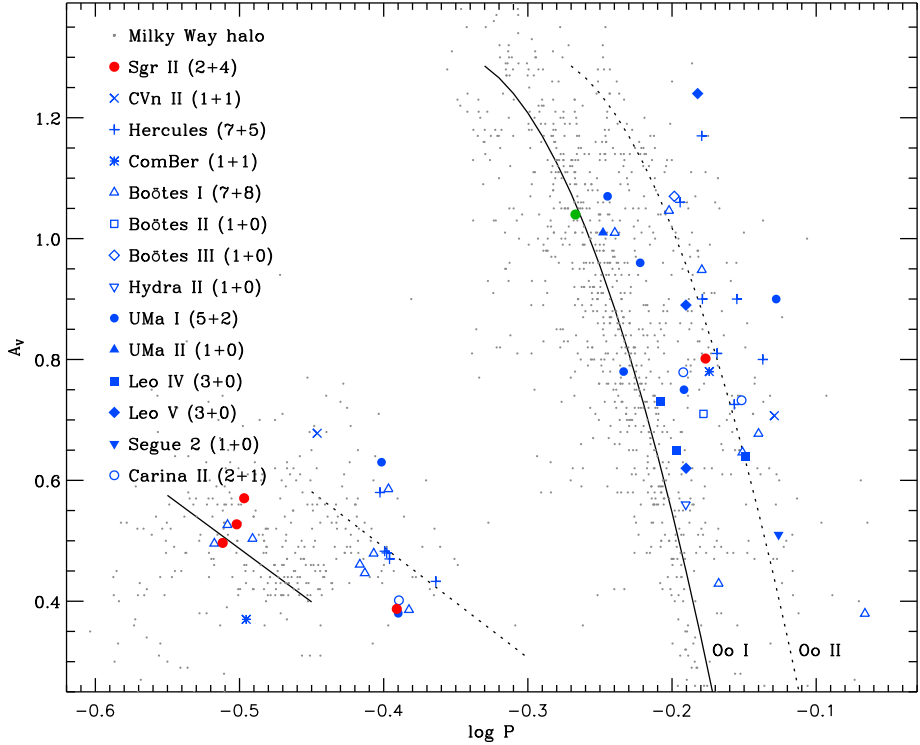}
\caption{
Period-amplitude diagram of the RR Lyrae stars in Sgr~II (filled red circles), other 13 UFDs (blue symbols), and the MW halo (gray dots).
The probable field star V4 is presented in the filled green circle.
The solid and dotted lines are the mean distribution of RR Lyrae stars in the MW GCs that belong to Oo~I and Oo~II groups,
respectively \citep{zor10,cac05}. Numbers in parentheses indicate the numbers of $ab$- and $c$-type RR Lyrae stars in each UFDs,
separated by plus signs. This diagram is an update of Figure~10 in \citet{viv16} and Figure~6 in \citetalias{joo18}.
The RR Lyrae data were taken from \citet{zin14} for the MW halo, \citet{gre08} for Canes~Venatici~II (CVn~II),
\citet{mus12} and \citet{gar18} for Hercules, \citet{mus09} for Coma~Berenices (ComBer), \citet{sie06} for Bo\"otes~I,
\citet{viv16} for Hydra~II, \citet{gar13} for Ursa~Major~I (UMa~I),
\citet{mor09} for Leo~IV, \citet{med17} for Leo~V, \citet{boe13} for Segue~II, and \citet{tor18} for Carina~II.
In the case of Bo\"otes~II, Bo\"otes~III \citep{ses14}, and UMa~II \citep{dal12}, we used the data revised by \citet{viv16},
and the $A_B$ values in Bo\"otes~I were converted to $A_V$ using equations in \citet{dor99} following \citet{viv16}.
The SDSS $g$-band amplitudes, $A_g$, of RR Lyrae variables in Hercules \citep{gar18}, Leo~V, and Carina~II,
were transformed to $A_V$, adopting the equation, $V = g - 0.59\,(g-r) - 0.01$, in \citet[see their Table~1]{jes05}, where for Leo~V
we assumed $(g-r) = 0.15$ as the mean color of RRab stars without the SDSS $r$-band magnitude \citepalias[see][]{joo18}.
\label{fig4}}
\end{figure*}

Almost all UFDs (with $M_V > -7$) studied for RR Lyrae stars so far have Oo-int or Oo~II properties, as shown in Figure~\ref{fig4}.
If these UFDs host a purely old and metal-poor population resulting in a substantially blue HB morphology,
this can be explained with the effect of evolution of RR Lyrae stars in the instability strip.
Because in that situation most RRab variables are highly evolved stars from the blue side of the instability strip,
they are expected to have longer periods than those at the zero-age HB phase \citep[e.g.,][]{lee90}.
The relatively long period of V3 in Sgr~II can also be explained by this evolution effect.

Another interesting property of Sgr~II is that, it is at the expected location of the trailing arm of the Sgr stream yet to be detected
behind the Sgr dwarf galaxy and therefore it might be originally a satellite of the Sgr galaxy but brought into the MW halo,
as suggested by \citet{lae15}.
While we could not find evidence for that hypothesis from RR Lyrae stars, if confirmed, it would mean that
the properties of UFD satellites belonging to the MW halo and the Sgr galaxy are similar in terms of their stellar populations (i.e., age and metallicity).
Combined with the expected association with the Magellanic Clouds of some UFDs in the southern hemisphere \citep[e.g.,][]{jet16,jer18},
this would further imply universal properties of UFDs in the universe as relics of the first galaxies
and/or surviving counterparts of the basic building blocks that merged and disrupted to form larger galaxies.

\acknowledgments

We thank the anonymous referee for a number of helpful comments and suggestions.
This research has made use of the KMTNet system operated by the Korea Astronomy and Space Science Institute (KASI)
and the data were obtained at one of three host sites, CTIO in Chile.
This work has made use of data from the European Space Agency (ESA) mission {\it Gaia} (\url{https://www.cosmos.esa.int/gaia}), processed by the {\it Gaia}
Data Processing and Analysis Consortium (DPAC, \url{https://www.cosmos.esa.int/web/gaia/dpac/consortium}). Funding for the DPAC
has been provided by national institutions, in particular the institutions participating in the {\it Gaia} Multilateral Agreement.
S.C.R was partially supported by the Basic Science Research Program through the NRF of Korea funded by the Ministry of Education (2018R1A2B2006445).
Support for this work was also provided by the NRF to the Center for Galaxy Evolution Research (2017R1A5A1070354).
H.Jerjen acknowledges the support of the Australian Research Council through Discovery Project DP150100862.
H.Jeong acknowledges support from the Basic Science Research Program through the National Research Foundation (NRF) of Korea,
funded by the Ministry of Education (NRF-2013R1A6A3A04064993).
This research was made possible through the use of the APASS, funded by the Robert Martin Ayers Sciences Fund.


\begin{deluxetable}{lcccccccccclccc}
\tabletypesize{\scriptsize}
\tablecaption{Pulsation Properties of RR Lyrae Stars in Sgr~II\label{tbl1}}
\tablehead{\\
\colhead{ID} & \colhead{R.A.} & \colhead{Dec.} & \colhead{Type} & \colhead{Period} & \colhead{$\langle B \rangle$\tablenotemark{i}}
& \colhead{$\langle V \rangle$\tablenotemark{i}} & \colhead{$N_B$\tablenotemark{ii}} & \colhead{$N_V$\tablenotemark{ii}}
& \colhead{$A_B$\tablenotemark{iii}} & \colhead{$A_V$\tablenotemark{iii}} & \multicolumn{4}{c}{$Gaia$ DR2}  \\  
\cline{12-15}
&&&&&&&&&&& \colhead{ID} & \colhead{Type} & \colhead{Period} & \colhead{$N_G$\tablenotemark{iv}}  \\ 
\colhead{} & \colhead{(2000)} & \colhead{(2000)} & \colhead{} & \colhead{(days)} & \colhead{(mag)} & \colhead{(mag)}
& \colhead{} & \colhead{} & \colhead{(mag)} & \colhead{(mag)} & \colhead{} & & \colhead{(days)} & \colhead{}
}
\startdata
 V1                   &  19:52:45.48  &  $-$22:02:27.88  &     c  &  0.3148  &  20.13  &  19.64  &  117  &  121  &  0.78  &  0.53  & 6864423852970123520\tablenotemark{v}  & -  & -  & -  \\
 V2                   &  19:52:56.75  &  $-$22:04:08.36  &     c  &  0.4065  &  20.07  &  19.68  &  116  &  121  &  0.56  &  0.39  & 6864422757758521984  &  c & 0.4065  &   32     \\
 V3                   &  19:52:44.40  &  $-$22:03:04.71  &    ab  &  0.6656  &  20.22  &  19.76  &  117  &  121  &  1.08  &  0.80  & 6864422993976659968  & ab & 0.6657  &   34     \\
 V4\tablenotemark{vi} &  19:52:13.13  &  $-$21:42:58.93  &    ab  &  0.5408  &  19.95  &  19.52  &  117  &  121  &  1.47  &  1.04  & 6865195302117134336  & ab & 0.5407  &   37     \\
 V5                   &  19:52:38.05  &  $-$22:03:30.45  &     c  &  0.3078  &  20.13  &  19.85  &  117  &  121  &  0.65  &  0.50  & 6864048408410304896  &  c & 0.3079  &   36     \\
 V6                   &  19:52:35.88  &  $-$22:01:59.81  &     c  &  0.3186  &  20.12  &  19.81  &  117  &  121  &  0.79  &  0.57  & 6864423994704275200  &  c & 0.3186  &   35     \\
\enddata
\tablenotetext{\rm i}{Intensity-weighted mean magnitude. V1 is slightly contaminated by bleeding from a saturated star.}
\tablenotetext{\rm ii}{Number of observations used in our light curve analysis.}
\tablenotetext{\rm iii}{Pulsation amplitude.}
\tablenotetext{\rm iv}{Number of data points for $G$-band in $Gaia$ DR2}
\tablenotetext{\rm v}{Not identified as a variable in $Gaia$ DR2.}
\tablenotemark{\rm vi}{Field?}
\end{deluxetable}


\clearpage

\clearpage

\end{document}